\documentclass[aps,a4paper,10pt]{revtex4}

\usepackage[dvips]{graphics,epsfig,rotating}
\usepackage{subfigure}
\usepackage{amstext}
\usepackage{amssymb}
\usepackage{amsmath}
\usepackage{float}
\usepackage{wrapfig}

\newcommand{\eg}{e.g.}
\newcommand{\ie}{i.e.}

\newcommand{\pri}{^{\prime}}

\newcommand{\I}{\text{i}}
\newcommand{\e}{\text{e}}

\begin{document}

\title{
       {\parbox[b]{\textwidth}{\rm \footnotesize \begin{flushright}
       K. H. Hughes (ed.) \\
       Dynamics of Open Quantum Systems \\
       \copyright \hspace{1ex} 2006, CCP6, Daresbury
        \end{flushright}}} \\
Dynamics of Quantum Dissipative Systems:\\
The Example of Quantum Brownian Motors}
\author{J. Peguiron}
\affiliation{Department of Physics and Astronomy, University of Basel, Klingelbergstrasse 82, 4056 Basel, Switzerland}

\date{\today}

\maketitle
\setcounter{page}{1}
\thispagestyle{plain}

\section{Introduction}

The description of dissipation in quantum mechanical systems is a question which has interested scientists since decades.
Brownian motors~\cite{AstPT02,APA02,ReiPR02}, \ie\ devices able to produce useful work out of thermal forces with the help of other unbiased forces, provide an ideal benchmark for the investigation of quantum dissipative systems~\cite{WeiBK99}, for two reasons. First, the interaction with a dissipative environment plays an essential role in the performance of Brownian motors. Second, dissipative tunneling enriches the dynamics of quantum Brownian motors with respect to their classical counterpart, inducing features such as current reversals as a function of temperature~\cite{ReiPRL97}. Experiments in the quantum regime have been reported in~\cite{LinSci99,MajPRL03}. Brownian motors belong to the family of ratchet systems, which exhibit directed transport in the presence of unbiased forces which need not necessarily have a thermal origin.

A simple model for a quantum ratchet system is provided by the Hamiltonian
\begin{equation}\label{eq-RatH}
H_{\text{R}}=p^2/2M+V(q)
\end{equation}
of a quantum particle of mass~$M$ in one-dimensional space. The asymmetric potential of periodicity~$L$ is completely characterized by the amplitudes~$V_{l}$ and
phases~$\varphi_{l}$ of its harmonics in the Fourier representation
\begin{equation}\label{eq-DefPot}
V(q)=\sum\nolimits_{l=1}^{\infty}{V_{l}\cos{\left(2\pi lq/L-\varphi_{l}\right)}}.
\end{equation}
In order to bring thermal fluctuations into the system, we let it interact with a bath of harmonic oscillators given by the standard Caldeira-Leggett Hamiltonian
\begin{equation}\label{Intro-BathH}
H_{\text{B}}=(1/2)\sum\nolimits_{\alpha=1}^{N_\text{O}}\left[p_\alpha^2/m_\alpha+m_\alpha\omega_\alpha^2\left(x_\alpha-c_\alpha q/m_\alpha\omega_\alpha^2\right)^2\right].
\end{equation}
The spectral density $J(\omega)=(\pi/2)\sum_{\alpha=1}^{N_\text{O}}(c_\alpha^2/m_\alpha\omega_\alpha){\delta(\omega-\omega_\alpha)}$ fully characterizes the bath. We consider an Ohmic bath with viscosity~$\eta$, \ie~$J(\omega)\sim\eta\omega$ at low
frequency~$\omega$. An additional contribution~$H_\text{ext}(t)=-F(t)q$ to the Hamiltonian accounts for an unbiased driving force~$F(t)$ breaking thermal equilibrium.

The quantity of interest is the stationary velocity~$v_\text{R}=\lim_{t\to\infty}\dot{q}(t)$. A nonzero particle velocity means that work is released, although all forces acting on the system are unbiased. We have developed two methods~\cite{GriPRL02,PegPRE05,PegCP06}, described in the next Sections, to evaluate the stationary velocity in the model presented above. 

\section{Quantum ratchets with few energy bands}

Our first method relies on a restriction to the low-energy dynamics of the ratchet system. Using Bloch theorem, the ratchet Hamiltonian~(\ref{eq-RatH}) can be diagonalized exactly, yielding energy eigenstates organized in an infinite number of bands. When the energies associated with the driving force and the temperature are much smaller than the lowest band gap, it is satisfactory to keep the few $N_\text{B}$~bands lying below the potential barrier only. After rotation to the position eingenbasis~$|m,j\rangle$ and use of the localized character of the retained low-energy states, the Hamiltonian takes the form of a multi-band tight-binding model, 
\begin{multline}\label{FBQR-HDVRII}
H_\text{TB}=\sum\nolimits_{j=-\infty}^\infty\Bigl[\sum\nolimits_{m=1}^{N_\text{B}}\varepsilon_m|m,j\rangle\langle m,j|+
\sum\nolimits_{m\ne m\pri=1}^{N_\text{B}}\Delta_{m\pri m}^\text{intra}|m\pri,j\rangle\langle m,j|\\
+\sum\nolimits_{m,m\pri=1}^{N_\text{B}}\left(\Delta_{m\pri m}^{\text{inter},f}|m\pri,j+1\rangle\langle m,j|+\Delta_{m\pri
m}^{\text{inter},b}|m\pri,j\rangle\langle m,j+1|\right)\Bigr],
\end{multline}
with on-site energies~$\varepsilon_m$. The interband couplings~$\Delta_{m\pri m}^\text{intra}$ allow for vibrational motion inside each potential well, whereas~$\Delta_{m\pri m}^\text{inter,f}$ and~$\Delta_{m\pri m}^\text{inter,b}$ allow for tunneling to the nearest neighboring wells.

The dissipative dynamics in this tight-binding system, driven by a non-adiabatic harmonic force~$F(t)=F\cos(\Omega t)$, 
can be investigated by means of real-time path integral techniques. One can derive a generalized master equation for the state populations~$P_{m,j}(t)$, in terms of the transition rates induced by the couplings. When the transition rates remain much smaller than the driving frequency~$\Omega$, one can average the dynamics over one driving period and obtain the averaged ratchet velocity at long times,
\begin{equation}
\bar{v}_\text{R}=L\sum\nolimits_{m,m\pri=1}^{N_\text{B}}p_m^\infty\left(\bar{\Gamma}_{m\pri m}^\text{inter,f}-\bar{\Gamma}_{m\pri
m}^\text{inter,b}\right).
\end{equation}
The asymptotic
population~$p_m^\infty=\lim_{t\to\infty}\sum_{j=-\infty}^\infty\bar{P}_{m,j}(t)$ of the band~$m$, is also a simple combination of the averaged transition rates~$\bar{\Gamma}_{m\pri m}^\text{intra}$, $\bar{\Gamma}_{m\pri m}^\text{inter,f}$, and~$\bar{\Gamma}_{m\pri m}^\text{inter,b}$.

The details of the evaluation, involving the computation of the transition rates up to second order in the tunneling couplings, can be found in~\cite{GriPRL02}. As a result, the ratchet velocity reveals a nonmonotonic dependence on the amplitude and frequency
of the driving force and on dissipation. Current inversions can be obtained by changing
any of these parameters. At the location of an inversion, the system thus experiences driving-induced localization.
 
The main drawback of this approach is that the validity regime for the truncation to a finite number of energy bands excludes the interesting limit of large driving amplitude or temperature, related to the classical limit. This has motivated the development of another approach presented in the next Section.

\section{Duality relation for quantum ratchets}

Our second method is based on a duality relation~\cite{FisPRB85} for a quantum dissipative system with a sinusoidal potential tilted by a time-independent (DC) force~$F$. We have generalized this relation to periodic potentials of arbitrary shape as given in~(\ref{eq-DefPot}). Here again, the evaluation of the particle velocity~$v_\text{DC}(F)$ at long times is based on real-time path integral techniques. One first performs an exact expansion of the propagator of the system in powers of the amplitudes~$V_{l}$ of the potential harmonics. After this transformation, the path integrals involve Gaussian integrals only, which can be evaluated. The resulting series expression, in powers of~$V_{l}$, can be related to the series expression, in powers of the couplings, for the particle velocity~$\tilde{v}_\text{DC}(F)$ in a single-band tight-binding model with non-nearest-neighbors tunneling couplings. This relation, valid a long times, reads
\begin{equation}\label{eq-DRv}
v_\text{DC}(F)=F/\eta-\tilde{v}_\text{DC}(F),
\end{equation}
where~$\eta$ denotes the viscosity in the original system. The Hamiltonian of the dual tight-binding system in which~$\tilde{v}_\text{DC}(F)$ has to be evaluated takes the form
\begin{equation}\label{eq-DefHTB}
H_{\text{TB}}=\sum\nolimits_{m=1}^{\infty}\sum\nolimits_{l=-\infty}^{\infty}\Bigl(\Delta_{m}|l+m\rangle\langle l|
+\Delta_{m}^{*}|l\rangle\langle l+m|\Bigr).
\end{equation}
A striking outcome of the calculation is the relation~$\Delta_l=(1/2)\ V_l\ \e^{\I\varphi_l}$ giving the coupling~$\Delta_{l}$ to the $l$th-order neighbor in terms of the amplitude~$V_l$ and phase~$\varphi_l$ of the $l$th harmonic of the original potential.
The tight-binding system~(\ref{eq-DefHTB}) is tilted by a time-independent force~$F$ as the original system. It is bilinearly coupled to a different bath of harmonic oscillators, characterized by the spectral density~$J_\text{TB}(\omega)=J(\omega)/[1+(\omega/\gamma)^2]$, thus Ohmic as~$J(\omega)$ with an additional Drude cutoff at the frequency~$\gamma=\eta/M$ associated with dissipation in the original system. The spatial
periodicity~$\tilde{L}$ of the tight-binding model is related to the periodicity~$L$ of the original system through~$\tilde{L}=L/\alpha$, with the dimensionless dissipation parameter~$\alpha=\eta L^2/2\pi\hbar$. As a consequence, the dissipation parameter is~$\tilde{\alpha}=1/\alpha$ in the tight-binding system. A regime of weak dissipation in one system is thus related to a regime of strong dissipation in the dual one. 

The particle velocity~$\tilde{v}_\text{DC}(F)$ in the tight-binding model can be evaluated from the transition rates between the tight-binding states, like in the problem described  in the previous Section. The duality relation~(\ref{eq-DRv}) can then be used to obtain the particle velocity~$v_\text{DC}(F)$ in the original  system tilted by the time-independent force~$F$. From this result, one can deduce the ratchet velocity~$v_\text{R}(F)=v_\text{DC}(F)+v_\text{DC}(-F)$ in a system driven by an unbiased bistable force switching between the values~$\pm F$, in the limit where the switching rate is much slower
than any other time scale of the system. 
Another interesting quantity is the velocity $v_\text{L}(K,F)=v_\text{DC}(K+F)+v_\text{DC}(K-F)$ in a ratchet system driven by a bistable force~$\pm F$ and subject to a~constant load force~$K$, showing the load characteristic of the ratchet system. If the sign of this velocity is opposite to the sign of the load force, it means that the ratchet system is able to yield work against the load, \eg\ to lift it.

\begin{figure}
\begin{center}
\includegraphics{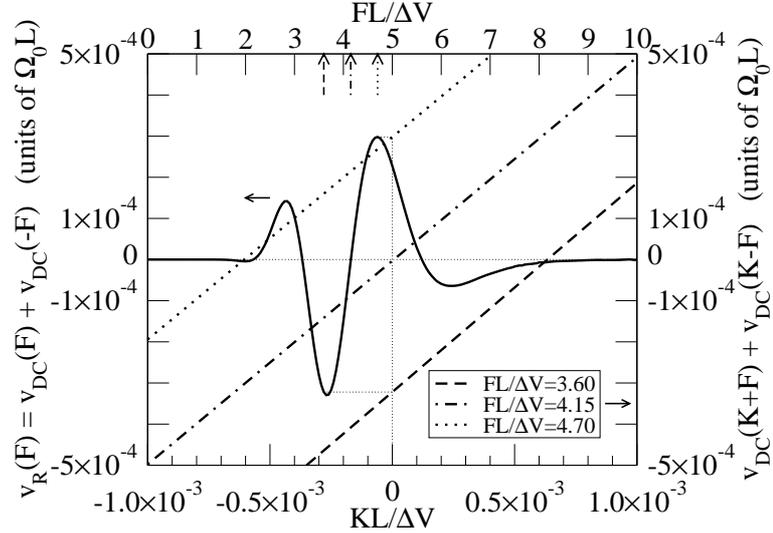}
\caption{\label{fig-loadcurve}Ratchet velocity (solid line, left and upper axes), and load characteristics (broken lines, right and lower axes) for a two-harmonics ratchet potential with amplitudes~$V_1=4\ V_2$, phases~$\varphi_2-2\varphi_1=-\pi/2$, and a spatial periodicity~$L$. The load force is denoted by~$K$ and the amplitude of the bistable driving force by~$F$. The vertical arrows show the driving amplitudes chosen for the computation of the load characteristics. The thin dotted lines are guides for the eyes. The temperature has been set to~$k_\text{B}T=0.1\ \Delta V$, in terms of the barrier height~$\Delta V=2.2\ V_1$. The viscosity~$\eta$ and the dynamical parameters of the system can be specified through the dissipation parameter~$\alpha=\eta L^2/2\pi\hbar$ and the ratio between the dissipation rate~$\gamma=\eta/M$ and the classical oscillation frequency in the untilted potential~$\Omega_0=2\pi\sqrt{V_1/ML^2}$. A regime of quantum dynamics and weak dissipation has been chosen with~$\alpha=0.2$ and~$\gamma=0.23\ \Omega_0$.}
\end{center}
\end{figure}

The ratchet velocity and load characteristics for a potential with two harmonics are shown in Fig.~\ref{fig-loadcurve}. The calculation follows the lines detailed in~\cite{PegPRE05,PegCP06}. The ratchet velocity (solid line, left and upper axes) shows several reversals as a function of the driving amplitude. It vanishes in the absence of driving. In this case the system is at equilibrium and no work can be extracted according to the Second Principle of Thermodynamics. The ratchet velocity also vanishes when the driving energy~$FL$ is much larger than the potential barrier~$\Delta V$, because the potential, which is the only source of spatial asymmetry in the system, is irrelevant in this regime. The figure also shows the load characteristics (broken lines, right and lower axes) as a function of the load force~$K$ for three different amplitudes of the bistable driving force. For~$FL=4.70\ \Delta V$~(dotted line), there is a range of negative load forces for which the velocity is positive. The ratchet system is able to work against the load in this regime. A range of positive load forces and negative velocities is obtained for~$FL=3.60\ \Delta V$~(dashed line). Finally, for~$FL=4.15\ \Delta V$~(dashed-dotted line), the load characteristic crosses the origin, meaning that the ratchet system is tuned into an idle regime.

An interesting extension of this work would be the evaluation of the diffusion coefficient and current noise in quantum ratchet systems.

\section{Acknowledgments}
The author is grateful to Milena Grifoni for a fruitful collaboration on the investigation of quantum ratchets, and to Peter H\"{a}nggi for interesting comments including the suggestion of Fig.~\ref{fig-loadcurve}. This work has been financially supported by the Swiss SNF and the NCCR Nanoscience.

\bibliographystyle{ccp6}
\bibliography{ccp6_qbm}

\end{document}